\documentclass[12pt]{article}
\usepackage{epsfig}

\newcommand{\ord}{{\cal O}}

\newcommand{\be}{\begin{equation}}
\newcommand{\ee}{\end{equation}}
\newcommand{\bea}{\begin{eqnarray*}}
\newcommand{\eea}{\end{eqnarray*}}
\newcommand{\ba}{\begin{eqnarray}}
\newcommand{\ea}{\end{eqnarray}}
\newcommand{\imlt}{\IM\lambda_t}
\newcommand{\RE}{{\rm Re}}
\newcommand{\IM}{{\rm Im}}
\newcommand{\Lms}{\Lambda_{\overline{\rm MS}}}

\def\epe{\varepsilon'/\varepsilon}

\begin{document}

\begin{titlepage}
\begin{flushright}
 TUM-HEP-420/01 \\
 UCL-IPT-01-08 \\
June 2001
\end{flushright}

\vspace*{15mm}

\begin{center}
{\LARGE \bf What is
the \boldmath{$(\varepsilon'/\varepsilon)_{\exp}$}
 Telling Us?}
\end{center}

\vspace*{5mm}

\begin{center}
{\large{\bf Andrzej J. Buras}}$^{\mbox a}$ {\large{\bf and
Jean-Marc G\'erard}}$^{\mbox b}$
\end{center}

\vspace*{10mm}

\hspace*{40mm}
\begin{minipage}{70mm}\baselineskip0.5cm
$^{\mbox a}$ Technische Universit\"at M\"unchen
\\
Physik Department
\\
D-85748 $\ \ $ Garching, Germany
\end{minipage}

\vspace*{10mm}

\hspace*{40mm}
\begin{minipage}{70mm}\baselineskip0.5cm
$^{\mbox b}$ Universit\'e catholique de Louvain
\\
Physics Department
\\
B-1348 $\ \ $ Louvain-la-Neuve, Belgium
\end{minipage}

\vspace*{10mm}

\centerline{\bf Abstract}
Nature might be kinder than previously thought as far
as $\epe$  is concerned. We show that the recently
obtained experimental  value for $\epe$ does not
require sizeable 1/N and isospin-breaking
corrections. We propose to display the theoretical
results for $\epe$ in a $(P^{1/2}, P^{3/2})$ plane 
in which the experimental  result is represented by a
$(\epe)_{\rm exp}$--path. This should allow to exhibit
transparently the role of 1/N and isospin-breaking
corrections in different calculations of
$\epe$. From now on theorists are allowed to walk only
along this $(\epe)_{\rm exp}$--path.

\end{titlepage}

\newpage

\section{Introduction}

The totally unexpected observation \cite{1} of a sizeable CP-violation
in the $K^0 - \bar K^0$ oscillations immediately triggered
theoretical speculations about a new superweak interaction \cite{2}
obeying the strict $|\Delta S | = 2$ selection rule. The large value of
the associated $\varepsilon$-parameter was then justified by the huge
amplification due to the tiny $K_L - K_S$ mass difference. Following this
rather simple picture, it was absolutely unlikely that CP-violation would
show up somewhere else in weak processes.


Almost exactly 37 years later, we know that superweak models have been
definitely ruled out by the new generation of high-precision experiments
on the $|\Delta S | = 1$ neutral $K$-decays. Indeed, the most recent
measurements of the associated
$\varepsilon'$-parameter that allows us to distinguish between
$\pi^+\pi^-$ and $\pi^0\pi^0$ final states in $K_L$ decays give
\begin{equation}\label{eprime1}
\RE(\varepsilon'/\varepsilon) =\left\{ \begin{array}{ll}
(15.3 \pm 2.6)\cdot 10^{-4} &{\rm (NA48)}~ \cite{NA48}~,\\
(20.7 \pm 2.8)\cdot 10^{-4} & {\rm (KTeV)}~\cite{KTEV}~.
\end{array} \right.
\end{equation}
Combining these results with earlier measurements by
NA31 collaboration at CERN  $ ((23.0\pm 6.5)\cdot 10^{-4})$
\cite{barr:93} and by the
E731 experiment at Fermilab $ ((7.4\pm 5.9)\cdot 10^{-4})$
\cite{gibbons:93} gives
the grand average
\be
\RE(\epe) = (17.2 \pm 1.8)\cdot 10^{-4}~.
\label{ga}
\ee


The Standard Model  for electroweak
and strong gauge interactions accomodates, in principle, both
$\varepsilon$ and $\varepsilon'$-parameters in terms of a single
CP-violating phase.
Rather early theoretical
attempts \cite{EGN} have
predicted $\epe$ between $10^{-2}$ and $10^{-4}$.
During the last decade a considerable progress in
calculating $\epe$ has been done by several groups.
These papers are reviewed in
\cite{REV} where all relevant references can be
found. The short distance contributions to $\epe$ are
fully under  control \cite{BJLW1}  but
the presence of considerable  long distance hadronic
uncertainties precludes a precise value of $\epe$ in the Standard
Model at present. Consequently, while theorists were
able to  predict the sign and the order of magnitude
of $\epe$, the range
\be\label{2}
(\varepsilon'/\varepsilon)_{\rm th} = (5 \ \mbox{to} \
30)\cdot 10^{-4}
\ee
shows that the present status of $(\epe)_{\rm th}$
cannot match the  experimental one.

Though really expected this time, the non-vanishing value of a second
CP-violating parameter has once again been determined
by our experimental colleagues.  However, one should
not forget the tremendous efforts made by theorists to
calculate $\epe$ in the Cabibbo-Kobayashi-Maskawa
paradigm \cite{CKM} of the Standard
Model. Simultaneously, one should not give up the
hope that one day theorists will be able to calculate
$\epe$ precisely. It is therefore important to have a
transparent presentation  of different theoretical
estimates of
$\epe$ in order to be able to  identify the patterns
of various contributions. On the other hand, having
for the first time the definite precise number for
$(\epe)_{\rm exp}$ it is crucial to learn what Nature is
trying to tell us about theory. In this note, we
intend to make  first steps in both directions.

\section{Basic Formulae}
The standard parametrization for the hadronic $K$-decays into two pions :
\ba
A(K^0 \to \pi^+\pi^-) &=& A_0 e^{i\delta} + {1\over \sqrt{2}} A_2
\nonumber \\
A(K^0 \to \pi^0\pi^0) &=& A_0 e^{i\delta} - \sqrt{2} A_2 \\
A(K^+ \to \pi^+\pi^0) &=& {3\over 2} A_2 \nonumber
\ea
contains the necessary ingredients to produce non-vanishing asymmetries.
For illustration consider
\ba
a_{CP} &\equiv& {\Gamma (K^0 \to \pi^+\pi^-) - \Gamma (\bar K^0 \to
\pi^+\pi^-)\over \Gamma (K^0 \to \pi^+\pi^-) + \Gamma (\bar K^0 \to
\pi^+\pi^-)} \nonumber\\
&=& {\sqrt{2}\sin \delta \over(1+\sqrt{2} \omega \cos \delta +
\omega^2/2)} \mbox{Im} \ \left({A_2\over A_0}\right)
\ea
where
\be
\omega \equiv {\mbox{Re} \ A_2\over \mbox{Re} \ A_0}~.
\ee
In order that $a_{CP}$ is non-vanishing
 the two partial isospin amplitudes $A_0$ and
$A_2$ must have a relative CP-conserving phase
(extracted from $\pi\pi$ scattering) which turns out
to be roughly equal to the phase of the
$\varepsilon$-parameter :
\be\label{COIN}
\delta  \approx \phi_\varepsilon \approx \pi/4
\ee
and a relative CP-violating phase
\be
\mbox{Im} \left({A_2\over A_0}\right) \neq 0.
\ee
These phases are nicely factorized in the physical parameter measuring
direct CP-violation in hadronic $K$-decays
\be\label{8}
\varepsilon' = {i\over \sqrt{2}} e^{-i\delta} \ \mbox{Im} \
\left({A_2\over A_0}\right)
\ee
if one defines
\ba
\eta_{+-} &\equiv& {A(K_L \to \pi^+\pi^-)\over A(K_S \to \pi^+\pi^-)}
\equiv \varepsilon + {\varepsilon'\over 1+{\omega\over \sqrt{2}}
e^{-i\delta}} \nonumber \\
&&\\
\eta_{00} &\equiv& {A(K_L \to \pi^0\pi^0)\over A(K_S \to \pi^0\pi^0)}
\equiv \varepsilon - {2\varepsilon'\over 1-\sqrt{2}\omega
e^{-i\delta}}~.
\nonumber
\ea
This allows to measure $\RE(\epe)$ through

\begin{equation}\label{BASE}
\RE(\epe)=\frac{1}{6}(1-\frac{\omega}{\sqrt{2}}
\cos\delta)
\left(1-\left|{{\eta_{00}}\over{\eta_{+-}}}\right|^2\right)
\end{equation}
where we have kept the small $\ord(\omega)$ correction usually dropped
by experimentalists but kept by theorists in the evaluation of 
$\varepsilon'$ using (\ref{8}). 
Notice that the coincidence displayed in (\ref{COIN})
implies an almost real  $\epe$ so that, already at this level, Nature is
kind to
us.

In the Standard Model, CP-violation only arises from the arbitrary quark
mass matrices. A straightforward diagonalization shifts then the unique
physical phase into the Cabibbo-Kobayashi-Maskawa (CKM) unitary mixing
matrix $V$ associated with the $V-A$ hadronic charged current
\be
J^{ab}_\mu = \bar q^a \gamma_\mu (1-\gamma_5)q^b  \equiv (\bar q^a q^b).
\ee
In this physical basis, we therefore have to start
with the classical current-current $\Delta S = 1$
Hamiltonian
\ba
{\cal H}^{\Delta S = 1} &\div& \sum_{q=u,c,t} \lambda_q J^{sq}_\mu
J^\mu_{qd} \hspace*{10mm} (\lambda_q \equiv V^\ast_{qs} V_{qd}) \nonumber
\\ &=& \lambda_u [(\bar s u)(\bar u d)-(\bar s c) (\bar c d)]_{\Delta I =
1/2, 3/2} \nonumber \\
&+& \lambda_t [(\bar s t)(\bar t d) - (\bar s c) (\bar c d)]_{\Delta I =
1/2}
\ea
to estimate the $A_0$ and $A_2$ partial decay amplitudes.

The $\Delta I = 1/2,3/2$ current-current operator
involving only the light $u, d$ and $s$ quarks is
just proportional to $\lambda_u$. A tree-level
hadronization into $K$ and $\pi$ mesons fields would
therefore imply $A_0 = \sqrt{2} A_2$, i.e. a vanishing
$\varepsilon'$-parameter (see (\ref{8})). In other
words, a non-zero
$\varepsilon'$-parameter is a pure quantum-loop effect in the Standard
Model. Notice that these loop effects are also welcome to explain the
empirical $\Delta I = 1/2$ rule :
\be
\omega_{\exp} \approx {1\over 22} \ll {1\over \sqrt{2}}.
\ee

The quantum transmutation of the heavy $t\bar t$ and $c\bar c$ quark
paires into light $u\bar u$ and $d\bar d$ ones which, eventually,
hadronize into final pion states allows now the pure $\Delta I = 1/2$
current-current operator proportional to $\lambda_t$ to contribute to
the $\Delta S = 1$ $K$-decays. In the most convenient CKM phase
convention, we have
\be\label{13}
\mbox{Im}  \lambda_u = 0
\ee
such that CP-violation only appears in the $A_0$
partial amplitude as long as isospin is strictly
respected in the ``heavy-to-light" transmutation
process. But in the Standard Model, neutral
transmutations are possible through heavy quark
annihilations into gluons, $Z^0$ or photon that are
represented by the so--called penguin diagrams.
While the latter electroweak contributions obviously
break isospin symmetry,   the former may also do so
by producing first an off-shell iso-singlet mesonic
state  (mainly $\eta$ or $\eta'$) which then turns
into an iso-triplet pion. These isospin-breaking (IB) effects respectively
induced by the electric charge difference $\Delta e = e_u - e_d$
and the mass splitting $\Delta m = m_u - m_d$ between
the up and the down quarks are usually expected to
show up at the percent level in weak decays. However,
a CP-violating
$\Delta I = 3/2$ amplitude turns out to be enhanced
by the famous
$\Delta I = 1/2$ rule factor $\omega^{-1}$ since
\be
\IM \left(\frac{A_2}{A_0}\right)
= - {\omega\over \mbox{Re}A_0} (\mbox{Im} A_0 -
{1\over\omega}\ \mbox{Im} A_2).
\ee
 From these quite general considerations, one concludes that
\be
(\varepsilon'/\varepsilon)_{th} = \mbox{Im} \lambda_t~ [P^{1/2} -
{1\over \omega} P^{3/2}]
\ee
with $P^{1/2}$ and $P^{3/2}$, two separately measurable quantities
defined with respect to the CKM phase convention defined in (\ref{13}).
Formally, $P^{1/2}$ and $P^{3/2}$ are given in terms of short distance
Wilson coefficients $y_i$ and the corresponding hadronic matrix
elements as follows
\begin{eqnarray}
P^{1/2} & = & r \sum y_i \langle Q_i\rangle_0~,
\label{eq:P12} \\
P^{3/2} & = &{r}
\sum y_i \left[\langle Q_i\rangle_2^{\Delta e}
+  \omega^{\Delta m} \langle Q_i\rangle_0\right]
\label{eq:P32}
\end{eqnarray}
where $r$ is a numerical constant and
\begin{equation}
\omega^{\Delta m} =  \frac{(\IM
A_2)^{\Delta m}}{\IM A_0}~.
\label{eq:Omegaeta}
\end{equation}

\section{The \boldmath{$(\epe)_{exp}$}-Path}
\begin{figure}[hbt]
\vspace{0.10in}
\centerline{
\epsfysize=3.0in
\epsffile{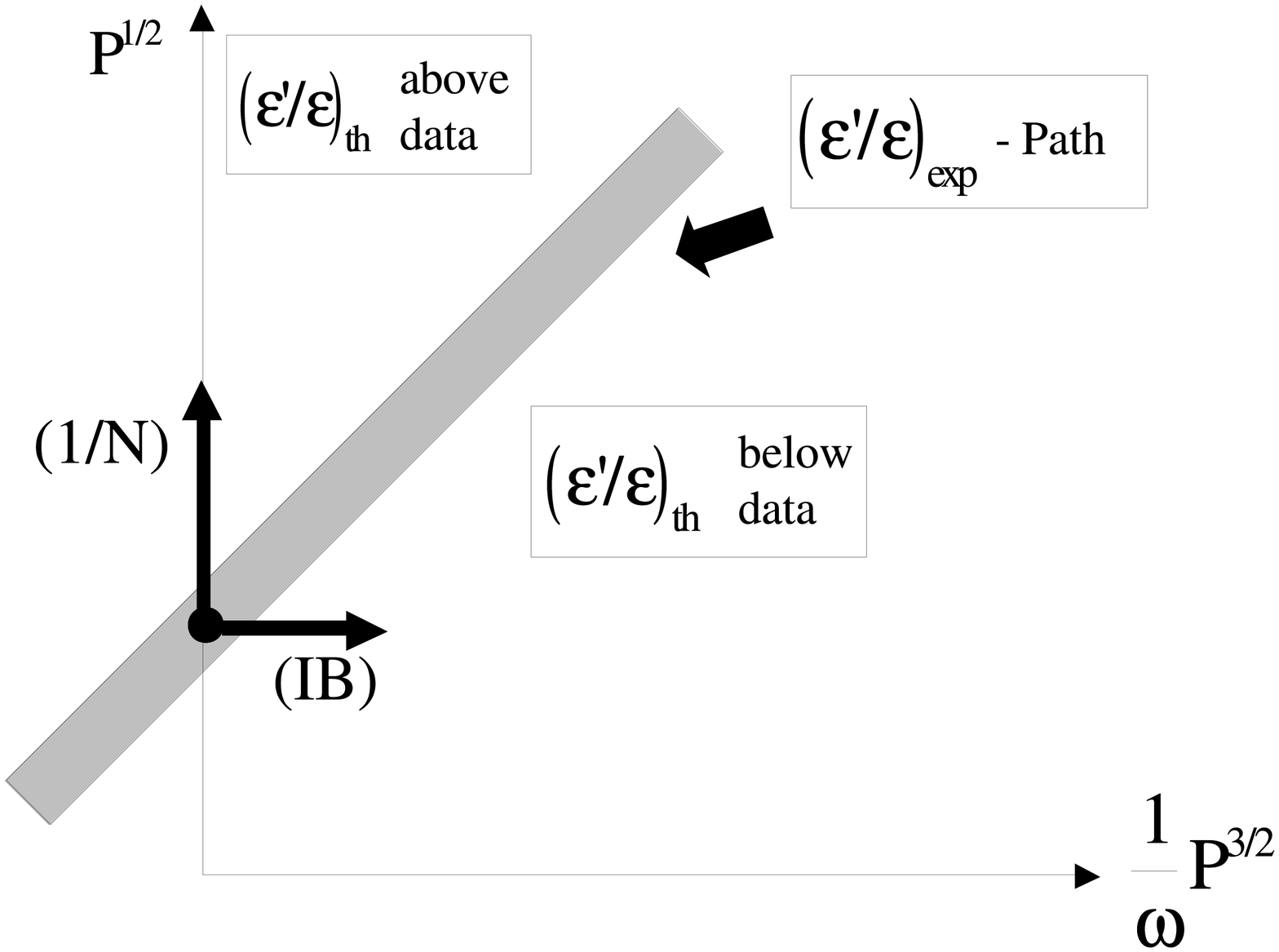}
}
\vspace{0.02in}
\caption{$(\epe)_{exp}$--path in the
$(P^{1/2},P^{3/2})$ plane.}
\label{BGPLOT}
\end{figure}

 Having all these formulae at hand, we can ask ourselves what the result in
(\ref{ga}) is telling us. The answer is simple. It allows us to walk
only along a
straight path in the $(P^{1/2},P^{3/2})$ plane, as illustrated in
Fig.\ref{BGPLOT}. The standard unitarity triangle analyses
\cite{C00} give typically
\be\label{TYP}
\IM \lambda_t = (1.2 \pm 0.2) 10^{-4}
\ee
and, combined with (\ref{ga}), already allow us to
draw a rather thin
$(\varepsilon'/\varepsilon)_{\exp}$-path in the $(P^{1/2},P^{3/2})$ plane
(see Fig.\ref{BGPLOT}). 
This path crosses the $P^{1/2}$--axis at 
$(P^{1/2})_0$ = $14.3 \pm 2.8$.

We are of course  still far away from such a
precise calculation of $P^{1/2}$ and $P^{3/2}$.
These two factors are dominated  by
the so-called strong $Q_6$ and electroweak $Q_8$ penguin
operators. The short-distance Wilson coefficients
$y_6$ and $y_8$ of these well-known density-density
operators are under excellent control \cite{BJLW1}. In particular,
the
$\Delta I = 3/2$ $Z^0$-exchange contribution to
$\epe$  exhibits a quadratic dependence on the top
quark mass which makes it to compete with the $\Delta
I = 1/2$ gluon-exchange one. Unfortunately, the
resulting destructive interference between $P^{1/2}$
and
$P^{3/2}$ strongly depends on the various hadronic
matrix elements. Long-distance effects are therefore
at the source of the large theoretical uncertainties
illustrated by (\ref{2}). Consequently, we  advocate to
adopt (temporarily) a different strategy to learn
something from the new precise measurements of
$\varepsilon'/\varepsilon$. The proposed exposition of $\epe$ in the
$(P^{1/2},P^{3/2})$ plane turns out to be useful in this context.

\section{A simple observation}

It is well-known that isospin-symmetry and large-$N$
limit represent two powerful approximations to study
long-distance hadronic physics. Here, these
well-defined approximations would allow us to neglect
$P^{3/2}$ and to express the hadronic matrix elements
of the surviving strong penguin operators  responsible for
$P^{1/2}$ in terms of measured form factors.
Earlier attempts \cite{bardeen:87} to go beyond such
a zero-order approximation provided us already with
some insight about the sign of the $1/N$ and IB
corrections to
$\epe$. Recent works including further ${1/ N}$
\cite{Prades} and IB
\cite{ECKER99} corrections confirm their tendancy to
increase  $P^{1/2}$ and $P^{3/2}$ respectively.
We illustrate these generic trends
\be
(\varepsilon'/\varepsilon)_{\rm th} =
(\varepsilon'/\varepsilon)_0 \{ 1 + {\cal O}
(1/N)-{1\over \omega} {\cal O} (IB)\}.
\ee
as $(1/N)$ and (IB) arrows in Fig.~\ref{BGPLOT}. 
A systematic calculation of all 1/N and IB corrections is not yet 
available, but a direct comparison
between the
measured value
$(\varepsilon'/\varepsilon)_{\exp}$ and the zero-order approximation
$(\varepsilon'/\varepsilon)_0$ should  already tell us something
about their magnitudes within the Standard Model. 
Indeed, if the experimental value quoted in
(\ref{ga}) is larger than the zero-order theoretical approximation, one
 needs $1/N$ corrections along the
$P^{1/2}$ axis :
\be
(\varepsilon'/\varepsilon)_{\exp} > (\varepsilon'/\varepsilon)_0
\Rightarrow 1/N \ \mbox{corrections}.
\ee
On the other hand, an experimental value smaller than the zero-order
approximation would be an indication for sizeable IB corrections along
the $P^{3/2}$ axis :
\be
(\varepsilon'/\varepsilon)_{\exp} < (\varepsilon'/\varepsilon)_0
\Rightarrow IB \ \mbox{corrections}.
\ee
And here comes the surprise ! It turns out that
$(\varepsilon'/\varepsilon)_0$ lies on the
$(\varepsilon'/\varepsilon)_{\exp}$-path in
Fig.~\ref{BGPLOT}. It is the crossing of this
path with the $P^{1/2}$ axis.

Indeed $(\varepsilon'/\varepsilon)_0$ can easily be estimated.
In the large-N limit, the non-perturbative parameter
$\hat B_K$  relevant for the usual analysis of the
unitarity triangle equals
$3/4$ \cite{MBW}. This implies
\be
\IM\lambda_t=(1.24 \pm 0.06)\cdot 10^{-4}
\ee
to be compared with (\ref{TYP}) that uses $\hat B_K=0.85\pm 0.15$.
Moreover, in the large-N limit the hadronic
matrix element of the strong penguin density-density
operator $Q_6$ factorizes $(B_6=1)$. A simple dependence on the
inverse of the strange quark mass squared arises then
to cancel the scale dependence of $y_6$ \cite{BUGE}.
Taking the central values of the strange quark mass
$m_s(2 GeV) = 110~{\rm  MeV}$  and of the QCD coupling
$\alpha_s (M_Z) = 0.119$ relevant for $y_6$, we obtain
\be\label{BG0}
(\varepsilon'/\varepsilon)_0 = (17.4\pm 0.7) \ 10^{-4}
\ee
where the error results from the error in
$\IM\lambda_t$. 
In obtaining (\ref{BG0}) we have taken also into account the contribution
of the other ($Q_4$) surviving  QCD penguin operator in the large-N limit. 
 Without this contribution we would find 
$18.4\pm 0.7$, still within the $(\varepsilon'/\varepsilon)_{\exp}$-path.
Clearly, as $(\varepsilon'/\varepsilon)_0$ is roughly proportional 
to $(\Lms^{(4)})^{0.8}/m_s^2$ with $\Lms^{(4)}=340\pm 40~{\rm MeV}$ 
and $m_s(2 GeV) = (110\pm 20)~{\rm  MeV}$, improvements on these input
parameters are mandatory.

Although this rather intriguing
coincidence between (\ref{ga}) and (\ref{BG0}) seems to indicate small 
$1/N$ and IB corrections, one cannot rule out a somewhat
accidental conspiracy between sizeable corrections
canceling each other
\be
{\cal O} (1/N) - {1\over \omega} {\cal O} (IB) \approx 0~.
\ee
The latter equation describes the walking along
the $(\epe)_{\exp}$--path.

At this point, it is also worth noticing that CP-violation
in the simplest  extensions of the Standard Model, the
models with minimal flavour-violation,  might behave
just like an IB correction along the $P^{3/2}$ axis.
The reason is that the $Z^0$-penguin maximally
violates the decoupling theorem. Consequently, it
depends quadratically on the top quark mass and is
also quite sensitive to new physics
\cite{BS00}. If such is the
case, one will have a hard time to disentangle new
sources of CP-violation beyond the Standard Model
from ordinary IB corrections.

Finally the $(\epe)_{\exp}$--path can be shifted vertically in the
$(P^{1/2}, P^{3/2})$ plane by new physics contributions to the
quantities used for the determination of $\imlt$ but this is
a different story.

\section{Conclusion}

Nature might be kinder than previously thought as far as
$\varepsilon'/\varepsilon$ is concerned. Indeed, present data do not
require sizeable $1/N$ and IB corrections. Improvements on the input
parameters $\alpha_s (M_Z)$ and $m_s$ leading to our estimate of
$(\varepsilon'/\varepsilon)_0$ are mandatory.
We have proposed to display the theoretical results
in a $(P^{1/2}, P^{3/2})$ plane in which the
experimental  result is represented by a
$(\epe)_{exp}$--path. This plot should allow to
exhibit transparently the role of 1/N and isospin-breaking corrections in
different theoretical results for
$\epe$.

\section*{Acknowledgments}

J.-M. G. appreciates the kind hospitality of the
Max-Planck-Institute for Physics in Munich where this work has been
initiated.

\par\noindent
This work has been supported in part by the German Bundesministerium f\"ur
Bildung und Forschung under the contract 05HT1WOA3.

\vfill\eject

\end{document}